\documentclass[12pt]{article}
\renewcommand{\baselinestretch}{1.2}
\usepackage{lineno}
\usepackage{bbm}
\usepackage{subcaption}
\usepackage{ragged2e}
\usepackage{indentfirst}
\usepackage{amsmath}
\usepackage{amssymb}
\usepackage{amsfonts}
\usepackage{amscd}
\usepackage{amsbsy}
\usepackage{amsthm}
\usepackage{slashed}
\usepackage{bm}
\usepackage{latexsym}
\usepackage{graphicx,color} 
\usepackage[dvipsnames]{xcolor}
\usepackage[colorlinks]{hyperref}
\hypersetup{linkcolor=blue,citecolor=teal,urlcolor=blue}

\usepackage[symbol]{footmisc} 
\usepackage{float} 
\usepackage{cite}  
\usepackage{titlesec}

\usepackage{ulem} 
\usepackage{orcidlink} 

\titleformat*{\section}{\large\bfseries}
\titleformat*{\subsection}{\normalsize\bfseries}
\begin{document}

\begin{center}
\renewcommand*{\thefootnote}{\fnsymbol{footnote}} 
{\Large \bf
     A Supersymmetric Extension of Axionic Electrodynamics:\\\vspace{8pt} From Axions and Photons to Axinos and Photinos
}
\vskip 6mm

{\bf C. Roldán-Domínguez}\,\orcidlink{0009-0000-5069-3878} \hspace{-1.75mm} $^\textrm{a}$ 
\hspace{-2mm} \footnote{Corresponding Author: \ clararoldan4@gmail.com},
\,
{\bf H.~Belich}\,\orcidlink{0000-0002-8795-1735} \hspace{-1.75mm} $^\textrm{a}$,
\hspace{-.mm}
\,
{\bf W.~Spalenza}\,\orcidlink{0000-0001-9644-3938} \hspace{-1.75mm} $^\textrm{b,c}$,
\hspace{-.mm}
\\[0.2cm]
{\bf A.L.M.A. Nogueira}\,\orcidlink{0009-0004-0736-9270} \hspace{-1.75mm}~$^\textrm{d}$,
\hspace{0.mm}
\,
{\bf M. Reetz}\,\orcidlink{0009-0001-5516-434X} \hspace{-1.75mm} $^\textrm{b}$
\hspace{-.mm}
\,
and
\,
{\bf J.A. Helayël-Neto}\,\orcidlink{0000-0001-8310-518X} \hspace{-1.75mm} $^\textrm{b}$
\hspace{-.mm}
\vskip 6mm
{$^\textrm{a}$ Universidade Federal do Esp\'{i}rito Santo, 29075-910, Vit\'{o}ria, ES, Brazil}
\\[0.2cm]
{$^\textrm{b}$
Centro Brasileiro de Pesquisas Físicas,
 22290-180, Rio de Janeiro, RJ, Brazil
}
\\[0.2cm]
{$^\textrm{c}$ Instituto Federal do Esp\'{i}rito Santo, 29150-410, Cariacica, ES, Brazil}
\\[0.2cm]
{$^\textrm{d}$ Centro Federal de Educação Tecnológica C. S. da Fonseca, 20271-110,\\ Rio de janeiro, RJ, Brazil}

\end{center}
\vskip 2mm
\vskip 2mm

\begin{abstract}
	
\noindent
In this contribution, we investigate a supersymmetric effective extension of Axionic Electrodynamics by adopting the superspace/superfield approach.  Rather than focusing on UV-complete axion models, our goal is to analyze how supersymmetry modifies the dynamical and propagating sectors of axion electrodynamics at the effective-field-theory level. In terms of component fields, the resulting Lagrangian describes the interactions among the axion, the photon, and their respective supersymmetric partners, the axino, the saxion and the photino. Supersymmetry induces new fermionic self-interactions and a non-polynomial interaction involving the axino, the photino, and the saxion. An interesting aspect to be highlighted is the appearance of fermionic bilinear contributions involving the photino to both the permittivity and permeability tensors of the constitutive relations. We also pay special attention to the dispersion relations in both the bosonic and fermionic sectors, and analyze the effective masses of the different particles  in the presence of an external magnetic background, which induces   supersymmetry  breaking. Finally, with the help of numerical methods, we identify a class of axionic and electromagnetic-field configurations with interesting profiles of the magnetic field.

\vskip 3mm


\end{abstract}
 \setcounter{footnote}{0} 
\renewcommand*{\thefootnote}{\arabic{footnote}} 

\newpage
\section{Introduction}
\label{sec:level1}

In the 1970s, strong interactions faced an intriguing problem: unlike weak interactions, they did not exhibit violation of charge and parity symmetry (CP violation). The absence of CP-symmetry violation in Quantum Chromodynamics (QCD) was addressed by a mechanism proposed by the Italian physicist Roberto Peccei and the Australian physicist Helen Quinn \cite{Peceiquin,weinberg,wilck}. Based on studies involving the emergence of axial current \cite{adler, Bell:1969ts, Bardeen:1969md}, they proposed a new $U(1)$ symmetry that would be spontaneously broken, generating a pseudo-Nambu-Goldstone boson, referred to as Axion \cite{kim, SHIFMAN1980493,DINE1981199,Zhitnitsky:1980tq}. Axions and axion-like particles (ALPs) arise in effective theories designed to extend the Standard Model (SM) of Particle Physics, aiming to address open problems in the interface of Particle Physics and gravitational interactions \cite{masso, kalos}.  Axions actually play an important role in the attempt to justify dark matter, the strong CP-problem, Condensed-Matter Physics and a great deal of issues related to astrophysical phenomena.

Due to their potential lightness and lack of electric charge, axions are extremely difficult to detect. In the model presented here, we study the axion–photon interaction, which provides a particularly promising detection channel, as it enables the conversion of axions/ALPs into photons in the presence of strong magnetic or electric fields.  This mechanism opens up the possibility of detecting galactic dark matter axions \cite{sikivie} as well as solar axions \cite{labo,Moriyama:1995bz,Inoue:2002qy,Zioutas:2004hi}, and is the basis of several current experiments \cite{coupling1, admxcollaboration2025searchaxiondarkmatter, arcusa2025, Ballou_2015, giannotti2024statusperspectivesaxionsearches}. This mixing between the axion and the photon appears in extensions of Electrodynamic Theory like Axionic Electrodynamics \cite{Paix_o_2022}. This theory supports the possibility of investigating observable axion effects, which can be tested in laboratory experiments, astrophysical observations \cite{Dicus:1978fp, Vysotsky:1978dc, Turner:1990uz, Raffelt:1990yz,  Raffelt:1996wa,  Raffelt:1999tx, Yao:2006px}, and cosmological studies \cite{Casaki, MARSH20161, Chen:1994ch, mirizi,Di_Luzio_2020,Sikivie_2021,Workman:2022ynf}. More recently,  the contributions of  ALPs to  electroweak precision observables have also been studied in \cite{Aiko_2023}. Axions are also investigated in the framework of high-energy cosmic ray physics  \cite{Csaki:2003ef, Gorbunov:2001gc} and  models that describe possible physics beyond the SM  \cite{dienes, dilella,  Carvalho_2023}. 

 An interesting proposal for Axion Electrodynamics with quantum corrections incorporated is reported in the work of Ref. \cite{petrov}. This article presents an axion-photon interaction term in which the axion field, $a(x)$, is coupled to a charged fermion through the parity-violating  interaction term $ia\,\overline{\psi}\gamma_{5}\psi$. The resulting low-energy effective theory is described by the following Lagrangian for pseudoscalar fields interacting with photons:
\begin{eqnarray}
\mathcal{L}_{\rm eff} &=& 
-\frac{1}{4} F_{\mu\nu} F^{\mu\nu} 
+ \frac{1}{2} \partial_\mu a \, \partial^\mu a  - \frac{1}{2} m_a^2 a^2
- \frac{g^{a\gamma}}{4} \, a \, F_{\mu\nu} \widetilde{F}^{\mu\nu} \,,
\label{eq1}
\end{eqnarray}
where $m_a$ is the axion mass and $g^{a\gamma}$ is the axion-photon coupling constant, which has 
dimensions of the inverse of mass; $F_{\mu\nu}$ is the electromagnetic field
strength, and $\tilde{F}^{\mu\nu}=\frac{1}{2}\varepsilon^{\mu\nu\kappa\lambda}F_{\kappa\lambda}$ stands for its dual. Our focus in this paper is to construct a supersymmetric extension of the effective axion–photon theory outlined in (\ref{eq1}). 

Before presenting our study on the $N=1$ supersymmetric version of the Axionic Electrodynamics, we consider it is mandatory to mention that there is an ample literature
on the association between axions and SUSY. Back to 1983, Kim \cite{Kim1984} related the SUSY-breaking scale of locally supersymmetric GUTs to the $10^{11}$ -- $10^{12}$ GeV scale of
the anomalous $U(1)$ Peccei-Quinn symmetry breaking scale, and pursued an investigation of the saxino (presently dubbed saxion -- the scalar supersymmetric partner of the axion
and axino) decay. We also quote the study of axions in connection with spontaneously broken supersymmetric theories, by coupling the axion supermultiplet to a SUSY-
breaking sector \cite{HigakiKitano2012}. Axions have also been inspected in the framework of supersymmetric GUTs with an intermediate SUSY scale, and an axion decay constant is naturally set in the range $10^9$ -- $10^{11}$ GeV \cite{HallNomuraShirai2014}. SaxiGUT theories are formulated in the paper \cite{CoDEramoHall2016}, where both the scale of the Peccei-Quinn breaking and the scale of the gauge-couplings unification can be precisely defined. The authors also show that saxion condensates could have been generated during the inflation era. The article \cite{ChoiYanagida2022} addresses the issue of quality and high-quality axions in scenarios with SUSY. More recently, high-quality axions are shown to emerge from symmetries intrinsic to higher-forms in extra dimensions \cite{CraigKongsore2025}. Finally, we
point out the works in which axion dark matter and axion-like particles (ALPs) appear from supersymmetric models and from scenarios with softly broken supersymmetries
\cite{BaerBargerSenguptaZhang2025} \cite{Ghosh2026}. After this quick survey on the literature that focus on the connection between axions and SUSY, we should clarify that, in the present contribution, we follow a different path: we do not couple the axion and its partners to a SUSY-breaking sector. Our paper sets out to simply formulate an (exact) N=1-SUSY version of the Axionic Electrodynamics as described by
the effective Lagrangian of Eq.(\ref{eq1}). Nevertheless, we should anticipate that, though our formulation does not contemplate SUSY breaking from the very onset, in the attainment of the dispersion relations for the bosonic and fermionic particles of the supersymmetric Axionic Electrodynamics, we find that the introduction of a background magnetic field to inspect the linearized version of the model yields a SUSY breaking (this shall be
discussed in Subsection (\ref{subsection4.2})), which can be read in the SUSY transformation of the photino field. It is neither a spontaneous breaking stemming from the vacuum expectation value of some auxiliary field nor from soft/hard breaking terms communicated to the low-energy scale. The breaking actually takes place by virtue of the anisotropy represented by the presence of the constant and homogeneous magnetic field. Though experiments and phenomenology confirm that SUSY is broken in the low-energy regime we have
been testing, our purpose is to understand how Axionic Electrodynamics behave in a scenario in which SUSY is still manifest with its breaking being realized by the presence
of an external magnetic field.

In the broad spectrum of (the by-now hypothetical) supersymmetric particles, axions can naturally arise as dark matter candidates in many extensions of the SM
\cite{ABBOTT1983133,Preskill:1982cy,DINE1983137}, including scenarios based on supersymmetry, extra dimensions and superstring theories. In addition to the neutralinos \cite{Catena2014,Abdughani2019}, new particles such as photinos and axinos have also been proposed as viable dark matter candidates \cite{FARRAR1996188, Fischler2011, steffen2007supersymmetricdarkmattercandidates}. Therefore, besides a purely exploratory attitude, we also have a physical motivation for extending Axionic Electrodynamics to a supersymmetric scenario, in which interactions between the axion and the photino — both dark matter candidates — are present. For the sake of clarity, we outline below the structure of the paper. 

Our study is divided into six main Sections. In the first Section, we present the foundations of our model, describe the action, and justify the natural emergence of the
axion and its scalar  and fermionic partners. In Section II, we analyze the resulting supersymmetric action and focus on the minimum of the potential involving the scalar fields, which provide relevant information on the masses of the axion and its partners. Next, in Section III, we investigate the modified Maxwell equations derived from this supersymmetric Axionic Electrodynamic model. In this context, we identify
fermionic contributions to both the polarization and magnetization as a consequence of the fermions brought about by supersymmetry. This is a new aspect not present in the
literature of anisotropic media and nonlinear post-Maxwellian extensions of Electromagnetic Theory. In Section IV, we focus our attention on the linearized regime of the model, obtain the dispersion relations (DR) for both the fermionic and bosonic sectors and identify the corresponding effective masses. Within this framework, we also justify the possible generation of vortices arising from the modified Ampère-Maxwell equation. In Section V, we further investigate the bosonic sector in the presence of a background field. We present graphical representations of different scenarios and discuss the role of the axion in the generation of special  magnetic field configurations. Finally, in the final Section, we outline future perspectives motivated by specific interaction terms of our action, which we believe may yield relevant physical insights.

To close this introductory Section, let us establish some conventions adopted throughout the paper. We will  work in flat Minkowski spacetime, without considering the effects of gravity, and with  signature $(+\,-\,-\,-)$. We also  employ natural units, such that $\hbar = c = 1$. Regarding the spinorial structure, spinor indices are denoted by $\alpha$ and should not be confused  with Minkowski spacetime indices, nor with the scalar saxion $\alpha$ or the  pseudoscalar axion $\beta$. These conventions shall be consistently used in the subsequent Sections.

\section{A supersymmetric and dynamical model for axions and axion-like particles}

\label{III}
The model we present in this paper is an $N = 1$-supersymmetric Abelian model for the axion.  The superaction that describes our model is given by the sum of the well-established super-QED field strength action and the novel coupling that mixes the photon and the axion, now in a supersymmetric context:   
\begin{eqnarray}
\mathcal{S}_{\textrm{\tiny SQED+axion}} &=& S_{\textrm{\tiny SQED}} + S_{\textrm{\tiny axion-photon}}  + S_{\textrm{\tiny axion dynamics}}, \label{lb}
\end{eqnarray}

with
\begin{equation}
\mathcal{S}_{\textrm{\tiny SQED}} =\frac{1}{4} \int d^{4}x d^{2}\theta d^{2}\bar{\theta} \bigg(\delta^2(\bar{\theta})\,W^{\alpha}W_{\alpha}
+\delta^2(\theta)\,\bar{W}_{\dot{\alpha}}\bar{W}^{\dot{\alpha}}\bigg), 
\end{equation}
and 
\begin{eqnarray}
\mathcal{S}_{\textrm{\tiny axion-photon}} &=& g^{a\gamma} \int d^4x \, d^2\theta \, d^2\bar{\theta} \, 
\bigg( W^\alpha (D_\alpha V) \mathcal{A}  + \bar{W}_{\dot{\alpha}} (\bar{D}^{\dot{\alpha}} V) \bar{\mathcal{A}} \bigg), \label{lb2}
\end{eqnarray}
where $W_{\alpha}=-\frac{1}{4}\bar{D}^{2}D_\alpha V$ is the supersymmetric field strength and $\bar{W}_{\dot{\alpha}}=-\frac{1}{4} D^2 \bar{D}_{\dot{\alpha}} V$. Notice that $V$ is the so-called vector superfield, which is written in the Wess-Zumino gauge as given below:
\begin{equation}
V_{\textrm{\tiny WZ}}(x,\theta,\bar{\theta})=\theta\sigma^{\mu}\bar{\theta}A_{\mu}(x)+\theta^{2}\bar{\theta}\bar{\lambda}(x)+\bar{\theta}^{2}\theta\lambda (x)+\theta^{2}\bar{\theta}^{2}d(x).
\end{equation}
Recall that \( \sigma^\mu = ( \mathbbm{1}_{2\times2}, \boldsymbol{\sigma} ) \) , where \( \boldsymbol{\sigma} \) denotes the usual Pauli matrices. The fields $A_{\mu}(x)$, $\lambda(x)$, and $d(x)$ are called component fields.  Specifically, \(A_\mu(x)\) is the photon gauge field (a real massless vector boson), \(\lambda(x)\) is the photino (gaugino) field (the Weyl spin-\(\tfrac{1}{2}\) fermionic superpartner of  \(A_\mu(x)\)), and \(d(x)\) is a real auxiliary scalar field that is included in order to close the supersymmetry algebra off shell. Thus, we have
$D_{\alpha}$ and $\bar{D}^{\dot{\alpha}}$ as the covariant supersymmetric derivatives: $D_{\alpha} =\partial_{\alpha}+i \sigma_{\alpha \dot{\beta}}^\mu \bar{\theta}^{\dot{\beta}} \partial_\mu$,   $\bar{D}_{\dot{\alpha}}  =-\bar{\partial}_{\dot{\alpha}}-i \theta^{\beta} \sigma_{\beta 
\dot{\alpha}}^\mu \partial_\mu$\,\, \footnote{The notation here is that $\mu, \,\nu,\, \kappa,\cdots$ are taken to be vector Lorentz indices,  $i,j,k,\cdots$ are Euclidean space indices, and finally $\alpha, \,\beta,\, \gamma,\cdots$ are spinor Lorentz indices.}. In order to achieve the axion-photon coupling superaction (\ref{lb2}), we propose, inspired by the Carroll-Field-Jackiw  supersymmetric scenario\footnote{We reinterpret the Carroll–Field–Jackiw term by performing a simple integration by parts of a scalar gradient and then promoting this scalar to a dynamical field, reinterpreting it as the axion. It is precisely because this field is dynamical that Lorentz symmetry is no longer violated in our framework.} presented in \cite{Belich_2003}, to build up the supersymmetric coupling term by means of a chiral superfield, as it constitutes the most natural and, a priori, simplest multiplet extension of the axion field to generate such a coupling.

Following this procedure, we generate the desired coupling between the axion and the photon. The scalar axion-like chiral superfield $\mathcal{A}$, which encodes the axion, is given by
\begin{align}
\mathcal{A}(x, \theta, \bar{\theta}) &=
a(x)
+ i \theta \sigma^\mu \bar{\theta} \partial_\mu a(x)
- \frac{1}{4} \bar{\theta}^2 \theta^2 \Box a(x) \nonumber \\
&\quad + \sqrt{2} \theta \psi(x)
+ \frac{i}{\sqrt{2}} \theta^2 \bar{\theta} \bar{\sigma}^\mu \partial_\mu \psi(x)
+ \theta^2 H(x),
\label{axion}
\end{align}
and 
\begin{align}
\bar{\mathcal{A}}(x, \theta, \bar{\theta}) &=
a^*(x)
- i \theta \sigma^\mu \bar{\theta} \partial_\mu a^*(x)
- \frac{1}{4} \bar{\theta}^2 \theta^2 \Box a^*(x) \nonumber \\
&\quad + \sqrt{2} \bar{\theta}\bar{\psi}(x)
- \frac{i}{\sqrt{2}} \bar{\theta}^2 \partial_\mu \bar{\psi}(x)\bar{\sigma}^{\mu} \theta
+ \bar{\theta}^2 H^{*}(x)
\label{axioncomplejo}
\end{align}
where the axion-like field, $a(x)=\alpha(x)+i\beta(x)$, accommodates the physical (pseudoscalar) axion field. Later on, we shall justify the identification of $\beta$ with the pseudoscalar axion. On the other hand, the fields $\alpha$ and $\psi$ correspond, respectively, to the scalar and fermionic superpartners of the pseudoscalar axion, $\beta$. 
Following the supersymmetry convention, the fermionic field $\psi$ is specifically referred to as the axino, while $\alpha$, the scalar superpartner of the pseudoscalar axion, $\beta$, is dubbed saxion. $H$ is again the complex auxiliary  field present in the axion supermultiplet. 

To conclude, we introduce the kinetic term for the superfield that accommodates the axion.  This requirement is satisfied by considering the third term in (\ref{lb}) to be,
\begin{align}
\mathcal{S}_{\textrm{\tiny axion dynamics}} &=
\int d^{4}x\, d^{2}\theta\, d^{2}\bar{\theta}\,
\bigg(
\bar{\mathcal{A}}\mathcal{A}
+ \frac{1}{2}m_a \mathcal{A}^2\delta^2(\bar{\theta})
+ \frac{1}{3}f\mathcal{A}^3\delta^2(\bar{\theta})
\nonumber \\
&\quad
+ \frac{1}{2}m^*_a \bar{\mathcal{A}}^2\delta^2(\theta)
+ \frac{1}{3}f^*\bar{\mathcal{A}}^3\delta^2(\theta)
\bigg)
\label{axiondinamicaction}
\end{align}
where $m_a$ and $m_a^*$ are mass parameters, and $f$ and $f^*$ are self-interaction parameters.

We here call attention to the fact that the kinetic term for the axion superfield in (\ref{axiondinamicaction}) corresponds to the usual kinetic term for a chiral scalar supermultiplet, as in the Wess-Zumino model. Unlike the work by Higaki and Kitano in \cite{HigakiKitano2012}, we do not propose a Kähler potential for the axion superfield with the shift symmetry associated with the $U(1)_{PQ}$ symmetry introduced by the authors. The reason is that we are not actually considering an effective axion model, as in \cite{HigakiKitano2012}. As already emphasized above, we are focusing exclusively on the simplest supersymmetric version of Axion Electrodynamics.

\section{The component-field action and the particle masses}

After some calculations, replacing (\ref{axion}) and (\ref{axioncomplejo}) into the different terms of (\ref{axiondinamicaction}), and assuming $m_a$ and $f$ to be real, we arrive at the component-field action. The Lagrange equations of motion for the auxiliary fields read
\begin{align}
H^* &= -f a^2 - m_a a - g^{a\gamma} \lambda^2, \nonumber \\
H   &= -f a^{*2} - m_a a^* - g^{a\gamma} \bar{\lambda}^2, \nonumber \\
d   &= \frac{\sqrt{2} g^{a\gamma} (\lambda \psi + \bar{\lambda} \bar{\psi})}{2 + 4 g^{a\gamma} (a + a^*)}.
\end{align}
Then, in order to extract information on the particle that we shall identify with the axion, one should bear in mind that the component field $a(x)$ that we are dealing with is complex and the axion degrees of freedom are described by the real or the imaginary part of $a(x)$. With $a=\alpha+i\beta$, $a^*=\alpha-i\beta$, we obtain

\begin{eqnarray}
\mathcal{S}_{\textrm{\tiny SQED+axion}} =&&\, \int d^4x\, \Bigg\{ - \frac{1}{4}F^{\mu\nu}F_{\mu\nu}
+ \partial^\mu \alpha\, \partial_\mu \alpha + \partial^\mu \beta\, \partial_\mu \beta  + i\partial_\mu \bar{\psi}\, \bar{\sigma}^\mu \psi \nonumber- i\lambda\sigma^\mu \partial_\mu \bar{\lambda}\nonumber \\&& 
- (g^{a\gamma})^2 \lambda^2 \bar{\lambda}^2-\frac{m_a}{2} (\psi^2 + \bar{\psi}^2)  - m_a(\alpha^2 + \beta^2)(m_a+2f\alpha)  
- 2i g^{a\gamma}(\bar{\lambda}^2 - \lambda^2) f \alpha \beta \nonumber \\&& - g^{a\gamma}(\lambda^2 + \bar{\lambda}^2)[m_a \alpha  + f(\alpha^2 - \beta^2)] +m_ai\beta g^{a\gamma}(\lambda^2-\bar{\lambda}^2) - f^2(\alpha^2 + \beta^2)^2 \nonumber\\&&- f(\psi^2 + \bar{\psi}^2)\alpha
- i f (\psi^2 - \bar{\psi}^2)\beta  +g^{a\gamma} \Big[-2i(\lambda\sigma^\mu\partial_\mu\bar{\lambda}-\partial_\mu\lambda\sigma^\mu\bar{\lambda})\alpha \nonumber\\&& +2(\lambda\sigma^\mu\partial_\mu\bar{\lambda}+\partial_\mu\lambda\sigma^\mu\bar{\lambda})\beta + \alpha F^{\mu\nu} F_{\mu\nu} - \beta \tilde{F}^{\mu\nu} F_{\mu\nu}   +\sqrt{2}(\lambda \sigma^{\mu\nu} \psi + \bar{\lambda} \bar{\sigma}^{\mu\nu} \bar{\psi}) F_{\mu\nu} 
\Big]  \nonumber\\&&-\frac{(g^{a\gamma})^2(\lambda\psi + \bar{\lambda}\bar{\psi})^2}{1 + 4g^{a\gamma} \alpha} 
\Bigg\}\,.\label{action_susy_Axion}
\end{eqnarray}
We then identify $\beta$ with the proper (pseudoscalar) axion field, as this is the choice that reproduces the Peccei-Quinn term that couples the axion to the photon. It can be proven that $\tilde{F}^{\mu\nu} F_{\mu\nu}$ is a pseudoscalar. Therefore, $\beta(x)$ must be a pseudoscalar in order to have a scalar action. This identifies the real axion with $\beta(x)$. We shall omit the term \textit{pseudoscalar} and simply refer to $\beta$ as the axion field. Interestingly, the coupling term $\alpha F^{\mu\nu} F_{\mu\nu}$, which naturally arises from supersymmetry, was already introduced in \cite{Ballou_2015}. The last non-polynomial term in the action is not usual in QED. It appears precisely when one makes use of the field equation for the auxiliary field, $d$, and replaces it in the action. As a matter of fact, one could expand the denominator of the non-polynomial term in powers of $g^{a\gamma}\alpha(x)$ and circumvent the non-polynomial character of the interaction. This is based on the assumption that one is not compelled to take into account third-order terms in $g^{a\gamma}$, since, regardless of the origin of the axion, $g^{a\gamma}$ is very small (at least of  order $10^{-10}$ GeV ), in line with what is accepted in the literature (see Sec. 90 (Axions and Other Similar Particles) in \cite{navas}). We may say that we are considering the weak field regime of the saxion $\alpha(x)$ field.

Now, we believe it is more appropriate to rewrite this action  in terms of four-component Majorana spinors $\Psi$, $\Lambda$ in the Weyl representation,
\begin{eqnarray}
S_{\textrm{\tiny SQED+axion}} &=& \int d^4x\, \bigg\{ - \frac{1}{4}F^{\mu\nu}F_{\mu\nu}
+ \partial^\mu \alpha\, \partial_\mu \alpha+ \partial^\mu \beta\, \partial_\mu \beta + g^{a\gamma} \alpha F^{\mu\nu} F_{\mu\nu} - g^{a\gamma} \beta \tilde{F}^{\mu\nu} F_{\mu\nu}  \nonumber\\
&& -\frac{(g^{a\gamma})^2}{1 + 4 g^{a\gamma} \alpha} \left( \bar{\Lambda} \Psi \right)^2 -\frac{1}{2} \bar{\Psi} \left[ i\,\gamma^\mu \partial_\mu + \left(m_a   + 2 f\, \alpha - 2 i  f\,  \gamma^5  \beta \right) \right] \Psi \nonumber\\
&& - \frac{1}{2}\, \bar{\Lambda} \left[i\gamma^\mu \partial_\mu   + 2g^{a\gamma} m_a \left( \alpha  + i \,  \beta  \gamma^5\right) \right] \Lambda  -2ig^{a\gamma} (\bar{\Lambda}\gamma^\mu\partial_\mu\Lambda)\alpha  \nonumber\\
&&- g^{a\gamma} \bar{\Lambda} \left[ f (\alpha^2 - \beta^2) + 2i  f \alpha \beta  \gamma^5\right]  \Lambda - \frac{1}{2} (g^{a\gamma})^2 (\bar{\Lambda} \Lambda )^2\, +2g^{a\gamma} (\bar{\Lambda}\gamma^\mu\gamma_5\partial_\mu\Lambda)\beta   \nonumber\\
&&+ g^{a\gamma}\sqrt{2}\, \bar{\Lambda} \Sigma^{\mu\nu} \Psi\, F_{\mu\nu}  -V(\alpha,\beta)\bigg\},
\label{accion4comp}
\end{eqnarray}
where
\begin{equation}
\Psi = \begin{pmatrix}
\psi \\
\bar{\psi}
\end{pmatrix},  \quad
\Lambda = \begin{pmatrix}
\lambda \\
\bar{\lambda}
\end{pmatrix}, \quad \Sigma^{\mu\nu} = \frac{i}{2} [\gamma^\mu, \gamma^\nu],
\end{equation}
and 
\begin{equation}
V(\alpha,\beta)= m_a(\alpha^2 + \beta^2)(m_a+2f\alpha) +f^2(\alpha^2 + \beta^2)^2 , \label{potential}
\end{equation}
which is the potential in the axionic sector. Let us present the analysis of this potential. 

At first glance, we notice  that the term $2f\alpha$ in the potential explicitly breaks the U(1) global symmetry for the axionic-like field $a=\alpha+i\beta$. To understand how this is expressed in the vacuum, we study the minima of the scalar potential (\ref{potential}). We compute  analytically  that this potential has no critical points
with $\beta\neq 0$, and we use the mass matrix  to calculate the character of those critical points, arriving at two minima and one saddle point. The two minima are $(\langle \alpha \rangle_1, \langle \beta \rangle_1) = (0, 0)$ and $(\langle \alpha \rangle_2, \langle \beta \rangle_2) = \left(- m_a /f,\, 0\right)$.  We then confirm this result by performing a numerical-computational sweep of parameters $m_a$ and $f$ over all possibilities of the potential. Moreover, we demonstrate that,  for both  minima of the potential for the scalars, the axion (CP-odd) $\beta$-field  is zero, i.e., $\beta$ has a vanishing vacuum expectation value $\langle \beta \rangle = 0$. This means that CP-symmetry remains unbroken in the vacuum (we should have in mind that we are dealing with electromagnetism). In contrast, the nontrivial vacuum expectation values of $\alpha$ indicates that the potential breaks  $U(1)$ global symmetry. Given this, both the axino, $\Psi$, and  photino, $\Lambda$, could acquire effective masses through their bilinear couplings to the scalar field. To verify whether such a mechanism really works, one needs to expand the axionic field around the non-trivial minimum of the potential, i.e., we consider that $\alpha(x) = \langle \alpha \rangle + \tilde{\alpha}(x) = -m_a/f + \tilde{\alpha}(x) $, and $ \beta(x) = \langle \beta \rangle + \tilde{\beta}(x) = \tilde{\beta}(x) $ and substitute it in the action.  We arrive at
\begin{eqnarray}
S_{\textrm{\tiny SQED+axion}} &=&\int d^4x\, \Big[ 
- \frac{1}{4}F^{\mu\nu}F_{\mu\nu} 
+ \partial^\mu \tilde{\alpha} \, \partial_\mu \tilde{\alpha} 
 + \partial^\mu \tilde{\beta}\, \partial_\mu \tilde{\beta} + \frac{i}{2}\,\partial_\mu \bar{\Psi} \gamma^\mu  \Psi - \frac{i}{2}\, \bar{\Lambda} \gamma^\mu \partial_\mu \Lambda \nonumber\\
&&- \frac{1}{2}(\bar{\Lambda} \Lambda )^2\, (g^{a\gamma})^2 
- g^{a\gamma}\bar{\Lambda} \Lambda \left(-m_a\tilde{\alpha} + f (\tilde{\alpha}^2 - \tilde{\beta}^2 ) \right) - g^{a\gamma} \tilde{\beta} \tilde{F}^{\mu\nu} F_{\mu\nu}
 \nonumber\\
&&- ig^{a\gamma}\bar{\Lambda}\left( -m_a + 2 f \tilde{\alpha} \right) \tilde{\beta}  \gamma^5 \Lambda + 2i g^{a\gamma} \frac{m_a}{f} (\bar{\Lambda}\gamma^\mu\partial_\mu\Lambda)- \bar{\Psi}\left( -\frac{m_a}{2}  + f \tilde{\alpha} - i f  \gamma^5  \tilde{\beta} \right)\Psi\nonumber\\
&& 
- 2i g^{a\gamma} (\bar{\Lambda}\gamma^\mu\partial_\mu\Lambda) \tilde{\alpha}   - \frac{(g^{a\gamma})^2}{1 - 4 g^{a\gamma} \frac{m_a}{f} + 4 g^{a\gamma} \tilde{\alpha}} ( \bar{\Lambda} \Psi )^2 + 2 g^{a\gamma} (\bar{\Lambda}\gamma^\mu\gamma_5\partial_\mu\Lambda) \tilde{\beta}
 \nonumber\\
&& + g^{a\gamma} \left(-\frac{m_a}{f} + \tilde{\alpha} \right) F^{\mu\nu} F_{\mu\nu} + g^{a\gamma}\sqrt{2}\, \bar{\Lambda} \Sigma^{\mu\nu} \Psi\, F_{\mu\nu}  
- V(\tilde{\alpha},\tilde{\beta})
\Big],
\label{acciontilde}
\end{eqnarray}
where, now we have,
\begin{eqnarray}
V(\tilde{\alpha},\tilde{\beta}) = f^2 \tilde{\alpha}^4 - 2 f m_a \tilde{\alpha}^3 + m_a^2 \tilde{\alpha}^2 
+ 2 f^2 \tilde{\alpha}^2 \tilde{\beta}^2 - 2 f m_a \tilde{\alpha} \tilde{\beta}^2  + m_a^2 \tilde{\beta}^2 + f^2 \tilde{\beta}^4 .
\end{eqnarray}
~\\
Now, to extract the mass of the axino, $\psi$,  we look at the  terms in the Lagrangian which yield the Dirac equation for this field,
$\int d^4x\,\left[\frac{i}{2} \partial_{\mu}\bar{\Psi} \gamma^\mu \Psi-\bar{\Psi}\left(-\frac{m_a}{2}\right)\Psi \right]$, which leads to the Dirac equation $(i\gamma^\mu \partial_\mu - m_a)\Psi = 0$, with mass $m^{\text{eff}}_{\Psi}=m_a$.  For the axion mass, we focus on the terms
\begin{equation}
\int d^4x\,[( \partial^\mu \tilde{\alpha} \, \partial_\mu \tilde{\alpha}-m_a^2 \tilde{\alpha}^2)+ (\partial^\mu \tilde{\beta}\, \partial_\mu \tilde{\beta}  - m_a^2 \tilde{\beta}^2 )],
\end{equation}
from which we get that $m^{\text{eff}}_{\alpha}=m^{\text{eff}}_{\beta}=m^{\text{eff}}_{\Psi}=m_a$.  This result motivates two remarks: on the one hand, we observe that although the original potential 
is not symmetric in \(\alpha\) and \(\beta\),  the expansion around the minimum can produce quadratic terms with an effective symmetry between both of them, and, as a result, the same masses for \(\alpha\) and \(\beta\). On the other hand, the mass of the axion $\beta$ is equal to that of the axino $\psi$, which indicates that supersymmetry remains unbroken. We also prove this by computing that the potential is null at the two minima. Note that due to the specific form of our potential, the same result would be obtained by expanding around the trivial minimum without shifting the fields. The equal mass values for the fermion and the two scalars are a peculiar feature derived from the shape of our potential, which in its turn is imposed by supersymmetry. In addition, we verify that this symmetry breaking  mechanism does not give mass to the photino $\Lambda$.

\section{The Maxwell-like field equations of supersymmetric Axionic Electrodynamics} \label{Modified Maxwell Equations from a Supersymmetric Axion QED}

The Euler-Lagrange equations of the electromagnetic $A^\mu$-field in our action (\ref{accion4comp}) read,
\begin{align}
\label{EoM_bianchi_original}
&( 4\,g^{a\gamma}\alpha-1)\,\partial_\nu F^{\mu\nu}+ 4\,g^{a\gamma}\,(\partial_\nu \alpha)\,F^{\mu\nu}
-4\,g^{a\gamma}\,(\partial_\nu \beta)\,\tilde{F}^{\mu\nu}=-2\sqrt{2}g^{a\gamma}\partial_\nu(\bar{\Lambda}\Sigma^{\mu\nu}\Psi)
\end{align}
where the last fermionic term resembles an effective current. The Gauss law for the electric field and the Ampère-Maxwell law   are modified:
\begin{align}
&(4\,g^{a\gamma}\alpha-1)\,\nabla\cdot\boldsymbol{E}
+4g^{a\gamma}\,\nabla\alpha\cdot\boldsymbol{E}
-4g^{a\gamma}\,\nabla\beta\cdot\boldsymbol{B}
=-\nabla\cdot\boldsymbol{P}_{\rm s},
\label{gauss}
\\[0.2cm]
&( 4\,g^{a\gamma}\alpha-1)\big(\nabla\times\boldsymbol{B}-\boldsymbol{\dot{E}}\big)
+4g^{a\gamma}(\dot{\beta}\,\boldsymbol{B}-\dot{\alpha}\,\boldsymbol{E})+4g^{a\gamma}(\nabla\alpha\times\boldsymbol{B} 
-\nabla\beta\times\boldsymbol{E})
=\dot{\boldsymbol{P}}_{\rm s} +\nabla\times\boldsymbol{M}_{\rm s}
\label{ampere}
\end{align}
where $P_{\rm s}^i \equiv 2\,g^{a\gamma}\sqrt{2}\;\bar\Lambda\Sigma^{0i}\Psi$, $M_{k}^s \equiv \tfrac12\varepsilon_{kij}\tau^{ij}$, and $\tau^{ij} \equiv 2\,g^{a\gamma}\sqrt{2}\;\bar\Lambda\Sigma^{ji}\Psi$. The quantities $P^{i}_{s}$ and $M^{k}_{s}$ can be interpreted as contributions to the polarization and  magnetization of a spinorial origin. By virtue of the tensor structure of the fermionic bilinear $\partial_\nu(\bar{\Lambda}\Sigma^{\mu\nu}\Psi)$ in (\ref{EoM_bianchi_original}), its projection into the three-dimensional space in the derivation of Maxwell's equations (\ref{gauss}) and (\ref{ampere}) leads to an effective induced magnetization-polarization current, $\partial_{t}\boldsymbol{P}_{s} + \nabla \times \boldsymbol{M}_{s}$,
and to an effective induced polarization charge, $-\nabla \cdot \boldsymbol{P}_{s}$. Although we are considering our model in the vacuum (external sources are absent). this vacuum effectively behaves as a polarized and magnetized medium due to the presence of supersymmetric fermions. This is analogous to the behavior of photons propagating in a plasma. In a more compact form, by considering constitutive relations, equation (\ref{gauss}) can be re-expressed as
\begin{equation}
\nabla\cdot\boldsymbol{D}=0\,,
\label{gauss2}
\end{equation}
where $\boldsymbol{D}=\epsilon\,\boldsymbol{E}-\xi\,\boldsymbol{B}+\boldsymbol{P}_{\rm s}$ with
$\epsilon(x)= 4\, g^{a\gamma}\,\alpha(x) - 1$ and $\xi(x)=4g^{a\gamma}\,\beta(x)$. Making use of the definition of polarization and using Majorana fermion identities, we get
\begin{equation}
\boldsymbol{D}=\epsilon\,\boldsymbol{E}-\xi\,\boldsymbol{B}-\,2\sqrt{2}g^{a\gamma}\;X_{\alpha}^i\Lambda_\alpha\,,
\end{equation}
where $\bar\Lambda_\beta(\Sigma^{0i})_{\beta\alpha}\Psi_\alpha = \bar\Psi_\beta(\Sigma^{0i})_{\beta\alpha}\Lambda_\alpha = X_{\alpha}^i\Lambda_\alpha$. Here, $\boldsymbol{D}$ is introduced as an effective electric displacement field induced by the axion background: its constitutive equation depends explicitly on the scalar fields $\alpha$ and $\beta$.  In the same way, equation (\ref{ampere}) is given by 
\begin{equation}
\nabla\times\boldsymbol H - \dot{\boldsymbol D} = 8g^{a\gamma} \,\nabla\beta\times\boldsymbol E\,.
\label{ampere2}
\end{equation}
In this case, we identify a magnetic field strength that also depends on $\alpha$ and $\beta$: $\boldsymbol{H}=\epsilon\,\boldsymbol{B}+\xi\,\boldsymbol{E}-\boldsymbol{M}_{\rm s}$. Again, using the definition of the magnetization,
\begin{equation}
\boldsymbol{H}=\epsilon\,\boldsymbol{B}+\xi\,\boldsymbol{E}+\,g^{a\gamma}\sqrt{2}\;W_{\alpha}^{k}\Lambda_\alpha
\end{equation}
where $\varepsilon^{kij} \,\bar\Lambda\Sigma^{ji}\Psi = \varepsilon^{kij}\,Y_{\alpha}^{ij}\Lambda_\alpha = W_{\alpha}^{k}\Lambda_\alpha$. The quantity $\epsilon$ could be interpreted as an $x$-dependent effective permittivity/permeability factor induced by the saxion background field, $\alpha$. On the other hand, due to the presence of the axion field, $\beta$, the effective fields $\boldsymbol{D}$ and $\boldsymbol{H}$ depend on both $E$ and $B$. This means that $B$ can induce polarization, and $E$ can contribute to magnetization. This behavior, absent in the ordinary vacuum, reflects a mixing between the electric and magnetic sectors, corresponding to what we refer to as bianisotropy. We conclude that this nontrivial axion/axino supersymmetric background modifies Maxwell’s equations in a way formally equivalent to propagation in a
magneto–electric (bianisotropic) medium, similar to axion electrodynamics in topological insulators~\cite{Sekine_2021}.

\section{Dispersion relations in the bosonic and fermionic sectors}\label{4}

In this Section, we shall compute the DRs of the resulting linearized model in a background defined by a constant magnetic field. To achieve this, let us focus on the action given in (\ref{acciontilde}), where both axionic partners, $\alpha$ and $\beta$, represent physical fluctuations, as they  are expanded around the non-trivial minimum of the potential.  As our first move, we expand the electromagnetic field linearly around its background configuration and single out perturbations: $F^{\mu\nu} \rightarrow F^{\mu\nu}=f^{\mu\nu}+F_{\textrm{\tiny B}}^{\mu\nu}$. We take the background electric field to be zero ($F^{i0}_{\textrm{\tiny B}}=0$) and the background magnetic field to be constant ($F^{ij}_\textrm{\tiny B}=-\varepsilon^{ijk}\boldsymbol{B}_{\textrm{\tiny B}\,k}$).  After replacing this in the action (\ref{acciontilde}), while maintaining only quadratic terms\footnote{The linear term $g^{a\gamma} \tilde{\alpha}F_{\textrm{\tiny B}} ^{\mu\nu} F_{\textrm{\tiny B} \mu\nu}$  can be reabsorbed through a field redefinition: $\tilde{\alpha}=\hat{\alpha}+(g^{a\gamma}F_{\textrm{\tiny B}} ^{\mu\nu} F_{\textrm{\tiny B} \mu\nu})/2m_a^2$. This shift in $\tilde{\alpha}$ has been omitted for simplicity, since it is constant and therefore does not affect the subsequent analysis.}, we get: 
\begin{align}
  S_{\text{\tiny{SQED+axion}}} =&\int d^4x\, \Bigg[ 
-\left( g^{a\gamma} \frac{m_a}{f} +  \frac{1}{4}\right)f^{\mu\nu}f_{\mu\nu}  + \partial^\mu \tilde{\beta}\, \partial_\mu \tilde{\beta}+ \partial^\mu \tilde{\alpha} \, \partial_\mu \tilde{\alpha}  + \frac{1}{2}\,\bar{\Psi} (i\partial_\mu \gamma^\mu  +m_a ) \Psi \nonumber\\
&+ 2g^{a\gamma} \tilde{\alpha}F_{\textrm{\tiny B}}^{\mu\nu}f_{\mu\nu}+ \frac{i}{2}\, \bar{\Lambda} \gamma^\mu \partial_\mu \Lambda \left(4g^{a\gamma} \frac{m_a}{f}-1 \right) - 2g^{a\gamma} \tilde{\beta} \tilde{F}_\textrm{\tiny B}^{\mu\nu} f_{\mu\nu} + g^{a\gamma}\sqrt{2}\, \bar{\Lambda} \Sigma^{\mu\nu} \Psi\, F_{\textrm{\tiny B} \mu\nu}\nonumber\\
&-   m_a^2 \tilde{\alpha}^2 
  - m_a^2 \tilde{\beta}^2
\Bigg].   \label{ação-expandida}
\end{align}

The introduction of an external magnetic field provides a good tool to investigate the consequences of supersymmetrizing axionic electrodynamics. In particular, it  will allow us to analyze modified dispersion relations and effective masses of the bosonic and fermionic sectors. This type of configuration is common in astrophysical scenarios where axion-photon conversion processes, such as the Primakoff effect in the presence of strong magnetic fields, may take place.
 
\subsection{The bosonic sector}
First, we will focus on the bosonic sector of the theory. The  Euler-Lagrange field equations for the axionic $\tilde{\alpha}$ and $\tilde{\beta}$ fields are given by
\begin{align}
&(\Box+m_a^2)\tilde{\alpha}=g^{a\gamma}F_{\textrm{\tiny B}}^{\mu\nu}f_{\mu\nu},
\label{equcampoalfa}
\\[0.2cm]
&(\Box+m_a^2)\tilde{\beta}=-g^{a\gamma}\tilde{F}_\textrm{\tiny B} ^{\mu\nu}f_{\mu\nu}\,,
\label{equcampobeta}
\end{align}
whereas the electromagnetic field obeys the equation
\begin{equation}
\kappa\partial_\mu f^{\mu\nu}
= g^{a\gamma}\, (\partial_\mu \tilde{\alpha})\, F_{\textrm{\tiny B}}^{\mu\nu}
- g^{a\gamma}\, (\partial_\mu \tilde{\beta})\, \tilde F_{\textrm{\tiny B}}^{\mu\nu},
\label{Maxwell_cov}
\end{equation}
where
\begin{equation}
    \kappa=\left(\frac14 + g^{a\gamma}\frac{m_a}{f}\right). 
\end{equation}
Recall that $f$ is here the self-interaction parameter brought about in (\ref{axiondinamicaction}).
Writing Eqs.~\eqref{equcampoalfa}-\eqref{Maxwell_cov} in components, we get the axion and saxion wave equations, as well as the modified Maxwell equations:
\begin{align}
&(\Box+m_a^2)\tilde{\alpha}=2g^{a\gamma}\boldsymbol{B}_{\textrm{\tiny B}}\cdot\boldsymbol{b},
\label{eqalfa tilde}
\\[0.2cm]
&(\Box+m_a^2)\tilde{\beta}=2g^{a\gamma}\boldsymbol{B}_{\textrm{\tiny B}}\cdot\boldsymbol{e},
\label{eqbetatilde}
\\[0.2cm]
&{\nabla} \cdot \boldsymbol{e}
={-}\frac{g^{a\gamma}}{\kappa}\, \boldsymbol{B}_{\textrm{\tiny B}}\cdot {\nabla} \tilde{\beta},
\label{Gauss_mod}
\\[0.2cm]
&{\nabla} \times \boldsymbol{b} 
=   \partial_t \boldsymbol{e}{+} \frac{g^{a\gamma}}{\kappa}\, (\partial_t \tilde{\beta})\, \boldsymbol{B}_{\textrm{\tiny B}}
+ \frac{g^{a\gamma}}{\kappa}\, {\nabla} \tilde{\alpha} \times \boldsymbol{B}_{\textrm{\tiny B}} .
\label{Ampere_mod}
\end{align}
Since there are no magnetic monopoles in this scenario, the Bianchi identity, $\partial_\mu \tilde{f}^{\mu\nu}=0$, yields the homogeneous Maxwell equations:
\begin{equation}
{\nabla} \cdot \boldsymbol{b} = 0,
\qquad
{\nabla} \times \boldsymbol{e}= - \partial_t \boldsymbol{b}.
\label{bianchi}
\end{equation}
We observe that the axion field $\tilde{\beta}$ is responsible for the effective charge density in Eq.~\eqref{Gauss_mod}:
\begin{equation}
\rho_{\rm eff}={-}\frac{g^{a\gamma}}{\kappa}\boldsymbol{B}_{\textrm{\tiny B}}\cdot\nabla\tilde{\beta}.
\end{equation}
On the other hand, from the modified Ampère-Maxwell equation \eqref{Ampere_mod}, we identify effective current
contributions, induced by both axionic fields. In particular, the term $\nabla\tilde{\alpha}\times\boldsymbol{B}_{\textrm{\tiny B}}$ behaves as a transverse effective current that may induce a vortical structure.  Now, to derive the wave equations for the fields, we work in momentum space and adopt a plane-wave ansatz:
\begin{equation}
\begin{pmatrix}
\boldsymbol e(x) \\
\boldsymbol b(x) \\
\tilde\alpha(x) \\
\tilde\beta(x)
\end{pmatrix}
=
\begin{pmatrix}
\boldsymbol{e}_0\\
\boldsymbol{b}_0 \\
\tilde\alpha_0 \\
\tilde\beta_0
\end{pmatrix}
e^{i(\boldsymbol k\cdot \boldsymbol x - \omega t)} ,
\end{equation}where $\boldsymbol{k}$ and $\omega$ characterize the collective propagation of the mixed axion–photon eigenmodes, which propagate coherently as a single monochromatic excitation. Replacing this ansatz into Eqs.~\eqref{eqalfa tilde}--\eqref{bianchi} and performing the Fourier transform, we obtain the saxion  and axion wave equations:
\begin{align}
&(\omega^2- \boldsymbol{k}^2-m_a^2)\tilde{\alpha}_0=-2g^{a\gamma}\boldsymbol{B_\textrm{\tiny B}}\cdot\boldsymbol{b}_0,
\label{alfalineal}
\\[0.2cm]
&(\omega^2- \boldsymbol{k}^2-m_a^2)\tilde{\beta}_0={-}2g^{a\gamma}\boldsymbol{B_{\textrm{\tiny B}} }\cdot\boldsymbol{e_0},
\label{betalineal}
\end{align}
and the modified Maxwell equations
\begin{align}
&\boldsymbol{k} \cdot \boldsymbol{e}_0
={-} \frac{g^{a\gamma}}{\kappa}\, (\boldsymbol{B}_{\textrm{\tiny B}}\cdot\boldsymbol k\,)\,\tilde{\beta}_0,
\label{Momenta-Gauss-Electric}
\\[0.2cm]
&\boldsymbol{k} \times \boldsymbol{e}_0= \omega\, \boldsymbol{b}_0 ,
\label{Momenta-Faraday-Lenz}
\\[0.2cm]
&\boldsymbol{k} \cdot \boldsymbol{b}_0 = 0,
\label{Momenta-Gauss-Magnetic}
\\[0.2cm]
&\boldsymbol{k} \times \boldsymbol{b}_0
=   -\omega\, \boldsymbol{e}_0{-} \frac{g^{a\gamma}}{\kappa}\,  \,(\omega\, \boldsymbol{B}_{\textrm{\tiny B}})\,\tilde{\beta}_0
+ \frac{g^{a\gamma}}{\kappa}\, (\boldsymbol{k} \times \boldsymbol{B}_{\textrm{\tiny B}}) \,\tilde{\alpha}_0 .
\label{Momenta-Ampere-Maxwell}
\end{align} 

From this, we obtain the wave equation for the  electric field,
\begin{align}
    (\omega^2-\boldsymbol{k}^2)\,\boldsymbol{e}_0=\frac{g^{a\gamma}}{\kappa}\Big({-}\omega^2\,\boldsymbol{B}_{\textrm{\tiny B}}\, \tilde{\beta}_0+\omega\,(\boldsymbol{k}\times\boldsymbol{B}_{\textrm{\tiny B}})\,\tilde{\alpha}_0{+}(\boldsymbol{B}_{\textrm{\tiny B}}\cdot\boldsymbol{k})\,\boldsymbol{k}\,\tilde{\beta}_0\Big)
    \label{Electric-Wave-Equation}
\end{align}
and the magnetic field,
\begin{align}
    (\omega^2-\boldsymbol{k}^2)\,\boldsymbol{b}_0=\frac{g^{a\gamma}}{\kappa}\Big(-\boldsymbol{k}^2\boldsymbol{B}_{\textrm{\tiny B}}\, \tilde{\alpha}_0{-}\omega\,(\boldsymbol{k}\times\boldsymbol{B}_{\textrm{\tiny B}})\,\tilde{\beta}_0+(\boldsymbol{B}_{\textrm{\tiny B}}\cdot\boldsymbol{k})\,\boldsymbol{k}\,\tilde{\alpha}_0\Big).
    \label{Magnetic-Wave-Equation}
\end{align}With this set of equations in hand, we are able to obtain the DRs of the theory. In order to obtain the DRs for the saxion and axion fields, we  perform a dot product between $\boldsymbol{B}_{\textrm{\tiny B}}$ and the wave equations for the electric field \eqref{Electric-Wave-Equation} and the magnetic field \eqref{Magnetic-Wave-Equation}. Then, we can substitute the content of the saxion and  axion wave equations, \eqref{alfalineal} and \eqref{betalineal}, into those new equations, which leads us to
\begin{align}
&\Delta_{\tilde{\alpha}}\,\tilde{\alpha}_0=0,\\[0.2cm]&\Delta_{\tilde{\beta}}\,\tilde{\beta}_0=0,
\end{align}where
\begin{align}
\Delta_{\tilde{\alpha}}\equiv &\,\kappa\, (\omega^2-\boldsymbol{k}^2)(\omega^2-\boldsymbol{k}^2-m_a^2)+2({g^{a\gamma}})^2\,(\boldsymbol{B}_{\textrm{\tiny B}}\cdot\boldsymbol k)^2-2({g^{a\gamma}})^2\,\boldsymbol{k}^2\boldsymbol{B}_{\textrm{\tiny B}}^2,\\[0.2cm]
    \Delta _{\tilde{\beta}}\equiv&\,\kappa\, (\omega^2-\boldsymbol{k}^2)(\omega^2-\boldsymbol{k}^2-m_a^2)+2({g^{a\gamma}})^2\,(\boldsymbol{B}_{\textrm{\tiny B}}\cdot\boldsymbol k)^2-2({g^{a\gamma}})^2\,\omega^2\boldsymbol{B}_{\textrm{\tiny B}}^2,
\end{align}
with $\Delta_{\tilde{\alpha}}=0$ and $\Delta_{\tilde{\beta}}=0$ being the saxion and the axion DRs, respectively. 

These relations are very interesting; they will be  DRs for the mixing photon-saxion and for the mixing photon-axion.  By analyzing the wave equations for $\tilde{\alpha}_0$ and $\tilde{\beta}_0$, \eqref{alfalineal} and \eqref{betalineal}, we see that the propagating magnetic field $\boldsymbol{b}_0$ acts as a source for the equation of $\tilde{\alpha}_0$, whereas the propagating electric field $\boldsymbol{e}_0$ acts as a source for the equation of $\tilde{\beta}_0$. Meanwhile, both $\tilde{\alpha}_0$ and $\tilde{\beta}_0$ act as sources for the photon equations. What is interesting to note is that, although there is mixing of the photon with both fields, saxion and axion, there is no mixing between $\tilde{\alpha}_0$ and $\tilde{\beta}_0$, since the propagating electric and magnetic fields are orthogonal. We will verify that these DRs are associated with the mixed state of the fields by deriving  the photon DR, which can be obtained from either of the propagating fields, $\boldsymbol{e}_0$ or $\boldsymbol{b}_0$. In what follows, we work in terms of $\boldsymbol{e}_0$. To this end, we eliminate the propagating magnetic field from the Ampère–Maxwell equation \eqref{Momenta-Ampere-Maxwell} using the Faraday–Lenz equation \eqref{Momenta-Faraday-Lenz}. We also remove the contributions of $\tilde{\alpha}_0$ and $\tilde{\beta}_0$ via their respective wave equations \eqref{alfalineal} and \eqref{betalineal}. This procedure allows us to rewrite the modified Ampère's equation only in terms of the electric field $\boldsymbol{e}_0$,
\begin{align}&
 2(g^{a\gamma})^2\left[\boldsymbol{k}( \boldsymbol{B}_{\textrm{\tiny B}} \cdot \boldsymbol{k} )(\boldsymbol{B}_{\textrm{\tiny B}}\cdot\boldsymbol{e}_0) -\omega^2\boldsymbol{B}_{\textrm{\tiny B}}(\boldsymbol{B}_{\textrm{\tiny B}}\cdot\boldsymbol{e}_0)+(\boldsymbol{B}_{\textrm{\tiny B}}\cdot(\boldsymbol{k}\times\boldsymbol{e}_0))(\boldsymbol{k} \times \boldsymbol{B}_{\textrm{\tiny B}} )\right]\nonumber \\
&\quad+\kappa  \;(\omega^2-\boldsymbol{k}^2)( \omega^2- \boldsymbol{k}^2-m_a^2)\boldsymbol{e}_0
=0.
\label{ampere43}
\end{align}
Implementing index notation and taking advantage of standard Levi-Civita tensor identities, we write (\ref{ampere43}) in the compact form,
\begin{align}
M_{nj} \boldsymbol e_{0n}=0,
\end{align}
where
\begin{align}
M_{nj} =&\left[\kappa  ( \omega^2- \boldsymbol{k}^2-m_a^2)  \;( \omega^2 - \boldsymbol{k}^2) \;
+2(g^{a\gamma})^2 ((\boldsymbol{B}_{\textrm{\tiny B}}\cdot\boldsymbol{k})^2-\boldsymbol{B}_{\textrm{\tiny B}}^2\boldsymbol{k}^2)\right] \delta_{nj}\nonumber\\&-2(g^{a\gamma})^2( \omega^2 - \boldsymbol{k}^2  ){\boldsymbol B_{\textrm{\tiny B}}}_n{\boldsymbol B_{\textrm{\tiny B}}}_j- 2(g^{a\gamma})^2(\boldsymbol{B}_{\textrm{\tiny B}} \cdot \boldsymbol{k}){\boldsymbol B_{\textrm{\tiny B}}}_n \boldsymbol k_j +2(g^{a\gamma})^2\boldsymbol{B}_{\textrm{\tiny B}}^2 \,\boldsymbol k_n \boldsymbol k_j. 
\end{align}
Non-trivial solutions exist only if 
\begin{equation}
    \det (M_{nj}) = 0.
    \label{determinante}
\end{equation}
This leads to the DRs of the mixed modes.
\begin{align}
    &\kappa\, (\omega^2-\boldsymbol{k}^2)(\omega^2-\boldsymbol{k}^2-m_a^2)=0,\label{1}\\
    & \Delta_{\tilde{\beta}}=0,\label{2}\\
    & \Delta_{\tilde{\alpha}}=0.\label{3}
\end{align}
Solving the first DR, \eqref{1}, we obtain, for positive energy solutions,
\begin{align}
    \omega&=|\boldsymbol{k}|, \label{sinmixing}\\
\omega&=\sqrt{\boldsymbol{k}^2+m_a^2}\,.
    \label{sinmixing2}
\end{align} 
These solutions do not correspond to the physical
scenario we are contemplating, namely, the system axion-photon propagating in the presence of a constant  external magnetic field,  $\boldsymbol{B}_{\textrm{\tiny B}}$. Though higher derivatives are not involved in the field equations, by virtue of the axion-photon mixing, upon  splitting the axion and photon field equations, the dispersion relations exhibit higher powers of the frequency. Consequently,
 more solutions appear than expected for the actual physical system. In this context, solutions (\ref{sinmixing}) and (\ref{sinmixing2}) would arise for a photon and an axion without mixing and in the absence of an external magnetic field.  On the other hand, solutions of  (\ref{2}) and (\ref{3}) are the actual DRs for the axion and the photon mixed by the external magnetic field.  This is a general feature of fields free from higher derivatives but mixed in kinetic and/or mass terms. Upon separating the wave equations for each field individually, one obtains higher powers of the form  $\omega^{2N}$,  where N stands for the number of fields mixed with one another in the bilinear part of the action. This same reasoning will be reassessed later on, as we shall present and discuss the dispersion relations for the doublet axino-photino: solutions corresponding to a free axino and a free photino will arise along with actual solutions in the presence of the external magnetic field.
By solving \eqref{2}, one is led to
\begin{align}
    \omega&=\frac{1}{\sqrt{2\kappa}}\sqrt{2\kappa\,\boldsymbol{k}^2+2({g^{a\gamma}})^2\boldsymbol{B}_{\textrm{\tiny B}}^2+\kappa\,m_a^2\,\pm\sqrt{(2({g^{a\gamma}})^2\boldsymbol{B}_{\textrm{\tiny B}}^2+\kappa\,m_a^2)^2+8\kappa\,({g^{a\gamma}})^2\boldsymbol{B}_{\textrm{\tiny B}}^2\boldsymbol{k}^2\sin^2{\theta}}},
\end{align}
where $\theta$ is the angle between $\boldsymbol{B}_{\textrm{\tiny B}}$ and $\boldsymbol{k}$. Those solutions will be associated with the mixing between the photon and the  axion $\tilde{\beta}_0$. If we evaluate them in the rest frame, we can obtain individual contributions, 
\begin{align}
&{\omega_0}_+=\sqrt{\frac{2({g^{a\gamma}})^2\boldsymbol{B}_{\textrm{\tiny B}}^2+\kappa\,m_a^2}{\kappa}},\label{betaeqrestframe}\\
&{\omega_0}_-=0,
\end{align}
where the positive sign will give us an effective mass ${\omega_0}_+$ for the  axion, and the one with the negative sign will lead us to the photon contribution. 
Finally, for the last one \eqref{3}, it is found that
\begin{align}
    \omega&=\frac{1}{\sqrt{2\kappa}}\sqrt{2\kappa\,\boldsymbol{k}^2+\kappa\,m_a^2\,\pm\sqrt{\kappa^2\,m_a^4+8\kappa\,({g^{a\gamma}})^2\boldsymbol{B}_{\textrm{\tiny B}}^2\boldsymbol{k}^2\sin^2{\theta}}}\,.
\end{align}
This solution codifies the mixing between the photon and the saxion $\tilde{\alpha}_0$. Here, once again, we can identify individual contributions as we adopt the rest frame. In this case, we obtain
\begin{align}
    &{\omega_0}_+=m_a,\\
    &{\omega_0}_-=0,
\end{align}
where both results correspond to the usual outcomes for an axion, with the positive sign, and for the photon, with the negative sign.

Before going further to find the effective mass correction, let us notice that the axion effective mass  induced by an external magnetic field has been previously derived in the non-supersymmetric case, as presented in the papers \cite{Ter_as_2018,Sikivie_2021}. Here, we extend the corresponding calculation to a supersymmetric context, where the axion appears as a component of the supermultiplet that includes the saxion and the axino. We shall now try to shed light on equation \eqref{betaeqrestframe} and analyze the effective mass correction for the axion field generated by the
axion-photon coupling. That is,

\begin{align}
&{\omega_0}_+=\sqrt{m_a^2 + \Delta m_a^{2}},
\end{align}
where
\begin{equation}
     \Delta m_a^{2}=\frac{8 \boldsymbol{B}_{\textrm{\tiny B}}^2 f (g^{a\gamma})^2}{f + 4 g^{a\gamma} m_a}.
     \label{delta m}
\end{equation}

\begin{figure}[H]
    \centering
    \includegraphics[width=0.5\linewidth]{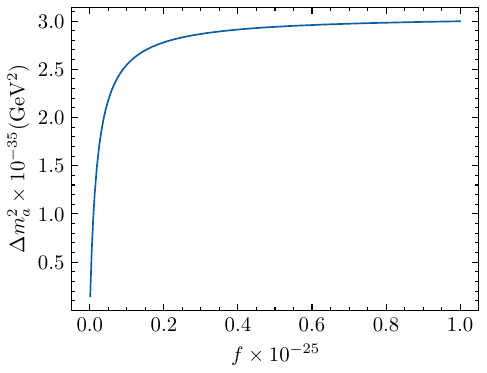}
    \caption{ Behavior of the mass-squared correction to  the  axion $\tilde{\beta}$ as a function of the self-interaction parameter $f$. 
In the limit $f \gg g^{a\gamma} m_a \approx 5\times10^{-28}$, the mass-squared correction approaches a constant value: 
$\Delta m_a^{2}\approx 8 \boldsymbol{B}_{\textrm{\tiny B}}^2 (g^{a\gamma})^2 \approx 3\times10^{-35}\,\mathrm{GeV}^2$. Conversely, in the limit $f \ll g^{a\gamma} m_a$ the mass-squared correction is negligible.  Therefore, in order to see how the parameter f influences the effective mass  we need to work in a regime where $f\approx g^{a\gamma}m_a$, more  exactly, with $f$ varying between $10^{-28}$ and $10^{-26}$. In this case, corrections to the mass-squared, although small,  are not negligible. In fact, the value adopted for the mass is $m_a^2=2.5\times10^{-13}$ eV$^2$, and using $f$ between $10^{-28}$ and $10^{-26}$ the correction to the mass-squared is of the order of $10^{-17}$ eV$^2$, that is, corrections at the fourth decimal place.}
    \label{betamasssquared}
\end{figure}

In order to compute the effective mass, we fix the relevant parameters.
We consider a magnetar background magnetic field $\boldsymbol{B}_{\textrm{\tiny B}} = 10^{10}\,\mathrm{T}\approx 195.5\times10^{-8}\,\mathrm{GeV}^2$. 
For the axion-photon coupling constant $g^{a\gamma}$ and the axion mass $m_a$, we adopt 
astrophysical bounds reported by the Particle Data Group \cite{PDG2023_axions}, 
namely $g^{a\gamma}= 10^{-12}\,\mathrm{GeV}^{-1}$ and $m_a = 5 \times 10^{-7}\,\mathrm{eV}$. 
With these values, Eq.~(\ref{delta m}) only depends on the parameter $f$, the self-interaction parameter coming from supersymmetry in Eq.~(\ref{axiondinamicaction}).
The result is shown graphically in Fig.~\ref{betamasssquared}.

\subsection{The fermionic sector}\label{subsection4.2}
The presence of an effective mass for the  axion could mean a breaking of supersymmetry  in this magnetic background scenario. To explore this, we will now turn our attention to the fermionic sector, examining its field equations and DRs. Considering the variational principle in the action \eqref{ação-expandida}, we obtain the field equations for the axino
\begin{align}
  (i\partial_\mu \gamma^\mu  +m_a )\Psi=g^{a\gamma}\sqrt{2} \Sigma^{\mu\nu}{F_{\textrm{\tiny B}}}_{\mu\nu}\Lambda ,
\end{align}and for the photino
\begin{align}
    i\kappa'\gamma^\mu \partial_\mu \Lambda =\frac{1}{4}g^{a\gamma}\sqrt{2}\Sigma^{\mu\nu}{F_{\textrm{\tiny B}}}_{\mu\nu}\Psi,
\end{align}where now, we have a new $\kappa'$  given by
\begin{align}
    \kappa'=\left(\frac{1}{4}-g^{a\gamma} \frac{m_a}{f}\right).
\end{align}
We rewrite the equations in momentum space representation, adopting a plane-wave ansatz
\begin{equation}
\begin{pmatrix}
\Psi(x) \\
\Lambda(x) 
\end{pmatrix}
=
\begin{pmatrix}
\Psi_0\\
\Lambda_0 
\end{pmatrix}
e^{i(\boldsymbol k\cdot \boldsymbol x - \omega t)} ,
\end{equation}where, once again, $\boldsymbol{k}$ and $\omega$ characterize the collective propagation of the mixed axino–photino eigenmodes, which propagate coherently as a single monochromatic excitation. Then, the equations become
\begin{align}
    &(\slashed{k} +m_a )\Psi_0=g^{a\gamma}\sqrt{2} \Sigma^{\mu\nu}{F_{\textrm{\tiny B}}}_{\mu\nu}\Lambda_0 ,\label{eq-axino}\\[0.2cm]
    &\kappa'\slashed{k}\Lambda_0 =\frac{1}{4}g^{a\gamma}\sqrt{2}\Sigma^{\mu\nu}{F_{\textrm{\tiny B}}}_{\mu\nu}\Psi_0.\label{eq-fotino}
\end{align}The equations are mixed, and to achieve our goal, it will be necessary to diagonalize them. In order to obtain decoupled equations, we perform a squaring procedure on Eqs. \eqref{eq-axino} and \eqref{eq-fotino}. Then,  we isolate contributions from the axino and the photino in those squared equations. Finally, we are able to write independent wave equations for both fields in the form
\begin{align}
    &\left(\kappa'k^2(k^2-m_a^2)-\frac{1}{2}({g^{a\gamma}})^2\slashed{k}\Sigma^{\alpha\beta}{F_{\textrm{\tiny B}}}_{\alpha\beta}\slashed{k}\Sigma^{\mu\nu}{F_{\textrm{\tiny B}}}_{\mu\nu}+\frac{1}{2}({g^{a\gamma}})^2m_a\Sigma^{\alpha\beta}{F_{\textrm{\tiny B}}}_{\alpha\beta}\slashed{k}\Sigma^{\mu\nu}{F_{\textrm{\tiny B}}}_{\mu\nu}\right)\Psi_0=0,\\
    &\left(\kappa'k^2(k^2-m_a^2)-\frac{1}{2}(g^{a\gamma})^2\slashed{k}\Sigma^{\alpha\beta}{F_{\textrm{\tiny B}}}_{\alpha\beta}\slashed{k}\Sigma^{\mu\nu}{F_{\textrm{\tiny B}}}_{\mu\nu}+\frac{1}{2}(g^{a\gamma})^2m_a\slashed{k}\Sigma^{\alpha\beta}{F_{\textrm{\tiny B}}}_{\alpha\beta}\Sigma^{\mu\nu}{F_{\textrm{\tiny B}}}_{\mu\nu}\right)\Lambda_0=0.
\end{align}
Although the above equations do not look the same, we can show, after some simplifications, that they lead us to the same DRs, namely,
\begin{align}
     &(\omega^2-\boldsymbol{k}^2)(\omega^2-\boldsymbol{k}^2-m_a^2)=0,\\[0.2cm]
    &\kappa'^2(\omega^2-\boldsymbol{k}^2)(\omega^2-\boldsymbol{k}^2-m_a^2)+\kappa'(g^{a\gamma})^2\left(8(\boldsymbol{B}_{\textrm{\tiny B}}\cdot\boldsymbol{k})^2-4\boldsymbol{B}_{\textrm{\tiny B}}^2\omega^2-4\boldsymbol{B}_{\textrm{\tiny B}}^2\boldsymbol{k}^2\right)+4(g^{a\gamma})^4\boldsymbol{B}_{\textrm{\tiny B}}^4=0.
\end{align}Their respective solutions are
\begin{align}
    &\omega=|\boldsymbol{k}|,\label{photino-livre}\\
    &\omega=\sqrt{\boldsymbol{k}^2+m_a^2}\label{axino-livre}\,,
\end{align}for the first DR, and
\begin{align}
\omega_\pm
&= \frac{1}{\sqrt{2\kappa'}} 
\Bigg\{
2\kappa'\,\boldsymbol{k}^2
+ \Big(4(g^{a\gamma})^2\boldsymbol{B}_{\textrm{\tiny B}}^2 + \kappa' m_a^2\Big)
\nonumber \\
&\quad \pm \sqrt{
\Big(4(g^{a\gamma})^2\boldsymbol{B}_{\textrm{\tiny B}}^2 + \kappa' m_a^2\Big)^2
- 16(g^{a\gamma})^4\boldsymbol{B}_{\textrm{\tiny B}}^4
+ 32\kappa'(g^{a\gamma})^2\boldsymbol{B}_{\textrm{\tiny B}}^2
\boldsymbol{k}^2 \sin^2\theta
}
\Bigg\}^{1/2}
\label{relacdispfotinoaxino}
\end{align}
for the second one. The solutions \eqref{photino-livre} and \eqref{axino-livre} are  expected when there  is no mixing between the particles. That diagonal mixing free case happens whenever one turns off the external magnetic field. However, since we are interested in exploring the scenario where a magnetic background prevails, we will focus on the solutions \eqref{relacdispfotinoaxino} that give effective masses for the photino and the axino. To compute those effective masses, we consider the rest frame configuration again. It follows  that
\begin{align}
   {\omega_0}_+=\frac{1}{\sqrt{2\kappa'}}\sqrt{(4(g^{a\gamma})^2\boldsymbol{B}_{\textrm{\tiny B}}^2+\kappa'\, m_a^2)+\sqrt{(4(g^{a\gamma})^2\boldsymbol{B}_{\textrm{\tiny B}}^2+\kappa'\, m_a^2)^2-16(g^{a\gamma})^4\boldsymbol{B}_{\textrm{\tiny B}}^4}},\\[0.2cm]{\omega_0}_-=\frac{1}{\sqrt{2\kappa'}}\sqrt{(4(g^{a\gamma})^2\boldsymbol{B}_{\textrm{\tiny B}}^2+\kappa'\, m_a^2)-\sqrt{(4(g^{a\gamma})^2\boldsymbol{B}_{\textrm{\tiny B}}^2+\kappa'\, m_a^2)^2-16(g^{a\gamma})^4\boldsymbol{B}_{\textrm{\tiny B}}^4}}.
\end{align} By setting $\boldsymbol{B}_{\textrm{\tiny B}}=0$ in those equations, we identify ${\omega_0}_+$ and ${\omega_0}_-$ as the axino and photino effective masses, respectively. As a result, we can write
\begin{align}
    {\omega_0}_+=\frac{1}{\sqrt{2}}\sqrt{m_a^2+(\Delta m_a^2)^+},
    \label{axinomassrecov}
\end{align}
and
\begin{align}
    {\omega_0}_-=\frac{1}{\sqrt{2}}\sqrt{m_a^2+(\Delta m_a^2)^-},
    \label{photinomassreco}
\end{align}where
\begin{align}
    (\Delta m_a^2)^\pm=\frac{16\boldsymbol{B}^2_{\textrm{\tiny B}}f(g^{a\gamma})^2}{f-4g^{a\gamma}m_a}\pm m_a\sqrt{m_a^2+\frac{32\boldsymbol{B}^2_{\textrm{\tiny B}}f(g^{a\gamma})^2}{f-4g^{a\gamma}m_a}}.\label{fermionic-mass-correction}
\end{align}

\begin{figure}[htbp]
    \centering
    \begin{subfigure}{0.48\linewidth}
        \centering
        \includegraphics[height=6 cm]{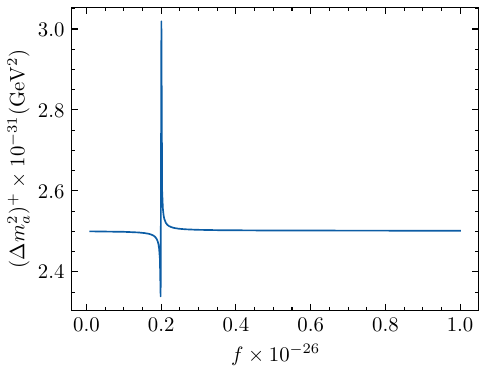}
        \caption{Curve for the axino field, $\Psi$.}
        \label{axino}
    \end{subfigure}
    \hfill
    \begin{subfigure}{0.48\linewidth}
        \centering
        \includegraphics[height=6 cm]{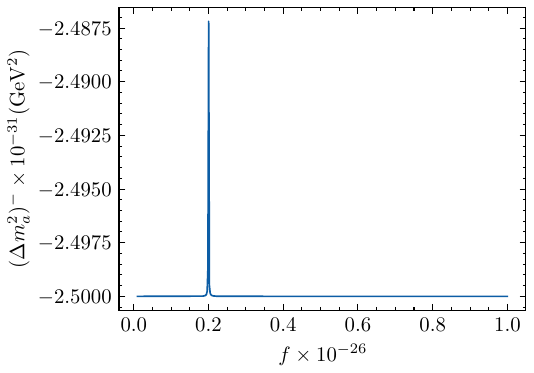}
        \caption{Curve for photino  field, $\Lambda$.}
        \label{photino}
    \end{subfigure}
    \caption{Behavior of the fermionic mass-squared corrections as a function of the self-interaction parameter $f$.}
    \label{fig:fermions}
\end{figure}

Analyzing the  limits  $f \to 0$ and $f\to \infty$ in Eq. \eqref{fermionic-mass-correction}, one can see that $(\Delta m_a^2)^\pm$ goes to $\pm m_a^2$. Taking these limits to Eqs.~(\ref{axinomassrecov}) and (\ref{photinomassreco}), we recover the onset axino and photino masses, $\omega_+=m_a$ and $\omega_-=0$ respectively, implying that the fermionic mass spectra does not change. When $f = 4 g^{a\gamma} m_a$, singularities arise that do not correspond to physical states, since this value eliminates the entire kinetic sector of the photino in the action \eqref{ação-expandida}. To analyze values of $f$ in the vicinity of the pole, one can assume the same magnetar background magnetic field $\boldsymbol{B}_{\textrm{\tiny B}} = 10^{10}\,\mathrm{T} \approx 195.5\times10^{-8}\,\mathrm{GeV}^2$ considered in the bosonic case, and use astrophysical bounds reported by the Particle Data Group \cite{PDG2023_axions}, where the axion-photon coupling constant $g^{a\gamma}$ and the axion mass $m_a$ are given by $g^{a\gamma}=10^{-12}\,\mathrm{GeV}^{-1}$ and $m_a = 5\times10^{-7}\,\mathrm{eV}$. In this case, we obtain from Eq. \eqref{fermionic-mass-correction} squared-mass corrections $(\Delta m_a^2)^\pm = \pm 2.5\times10^{-31}\,\mathrm{GeV}^2$. The behavior of these squared-mass corrections is shown in Fig.~\ref{axino} for the axino mass and in Fig.~\ref{photino} for the photino mass. From these plots one can see that near the pole, both mass corrections quickly assume constant values. 

From our previous developments, it can be readily checked that the bosonic and fermionic sectors do not share the same DRs; the degeneracy has been lifted. This points to a  breaking of supersymmetry. Although the original action in Eq. (9) is supersymmetric, the linearization around the (constant and homogeneous) background magnetic field, $\boldsymbol{B}_{\textrm{\tiny B}}$, induces a SUSY breakingdown. This can be explicitly verified by considering the behavior of the photino field under a SUSY transformation.
Once we expand the electromagnetic field around a background by  splitting the electromagnetic field strength into a sum of the background contribution and the propagating excitation, a non-trivial photino field variation emerges as a consequence of the SUSY transformation law, 
$\delta \lambda_\alpha= \frac{1}{2}(\sigma^{\mu\nu}\epsilon)_\alpha\, F_{\mu\nu}$, signaling a soft supersymmetry breaking. As far as the Lorentz symmetry is concerned, the background field introduces a space anisotropy in the vacuum, which can be thought of as a Lorentz-symmetry violation (LSV) from the particle point of view. In \cite{gauginoLSVrodrigo,Melo_2024}, it is illustrated how a LSV induced by a space-time anisotropy may yield a SUSY breaking identified in the (SUSY) transformation of a fermionic partner. This is a feature that we should comment on. As we have emphasized in the Introduction, we are interested in the investigation of aspects and peculiarities of Axionic Electrodynamics in connection with SUSY. What emerges from the linearized regime of the present model is that it undergoes a SUSY breaking whenever an external magnetic (or electric) field is switched on. This is tantamount to saying that the magnetic Primakoff effect in a supersymmetric context is accompanied by a SUSY breaking.

\section{Analyzing the bosonic sector in presence of an external homogeneous magnetic field}

\begin{figure}[H]
    \centering
\includegraphics[width=0.3\linewidth]{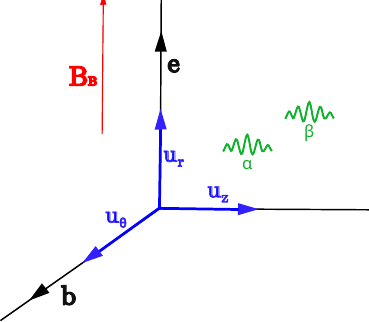}  \caption{Proposed setup in cylindrical coordinates. We consider a background magnetic field $\boldsymbol{B}_{\textrm{\tiny B}}$ aligned with an electric field $\boldsymbol{e}$ along the radial direction and orthogonal to the  magnetic field perturbation. The  fields $\alpha$ and $\beta$ represent the axionic excitations.}
  \label{fig:setup}
\end{figure}

As discussed in Sec.~\ref{Modified Maxwell Equations from a Supersymmetric Axion QED}, the presence of the axion field $\beta$ induces $E$–$B$ cross couplings in the electromagnetic fields $\boldsymbol{D}$ and $\boldsymbol{H}$. As a consequence, the model may exhibit nontrivial electromagnetic responses, such as the generation of charged electromagnetic field configurations, as we shall demonstrate in this Section. The interaction term $g^{a\gamma} \beta \tilde{F}^{\mu\nu} F_{\mu\nu}\sim 4 g^{a\gamma}\,\beta~{(\boldsymbol{E}\cdot\boldsymbol{B})}$ in the action (\ref{accion4comp}), which justifies the axion-photon interaction, is  a topological term whose physical effects become relevant in the presence of external fields. Moreover, the resulting theory may display features similar to phenomena in two spatial dimensions, such as the quantum Hall effect, whose effective description involves Chern–Simons terms. Motivated by these features, we propose a setup, schematically shown in Fig.~\ref{fig:setup}, in which a background magnetic field, in the presence of a dynamical electric field, interacts with  the axion $\beta$.  We do not specify the underlying physical system. The saxion $\alpha$ does not directly participate in this interaction, but it is still present in the theory, modifying the electromagnetic dynamics as derived in the previous Section. In order to investigate this scenario, we linearly expand the electromagnetic field around its background value and we single out perturbations, as in the previous Section, assuming a purely magnetic background field.   An interaction term of the form 
$g^{a\gamma}\beta(\boldsymbol{e}\cdot \boldsymbol{B}_{\textrm{\tiny B}})$
 emerges. Note that our goal here is to qualitatively analyze the behavior of the fields and potentials, rather than to make precise quantitative predictions.

We assume cylindrical symmetry inspired by the study of cosmic strings \cite{Vilenkin:2000jqa}, in which the fermionic fields can be consistently set to zero, allowing us to work only with the bosonic sector of the theory.  In this approximation, we study the behavior of the static electromagnetic four-potential $A^\mu = (V(r),A(r)\hat{e}_z)$.  From this ansatz, the fields read  $ \boldsymbol{e}=- V'(r)\hat{e}_r $ and $ \boldsymbol{b}=-A'(r)\hat{e}_\theta$. The  axionic  fields and the background magnetic field are assumed to share the same cylindrical symmetry, $\alpha=\alpha(r)$, $\beta = \beta(r)$, and $\boldsymbol{B}_{\textrm{\tiny B}} = B_{\textrm{\tiny B}}\hat{e}_r$. 
Under these assumptions, the equations of motion derived from the action (\ref{accion4comp}) take the form
\begin{eqnarray}
\alpha'' &=& -\frac{1}{r} \alpha' -  m^2 \alpha - 6fm\alpha^2 - 4f^2\alpha(\alpha^2 + \beta^2)+ g^{a\gamma}[e^2 - (b^2 + B_{\textrm{\tiny B}}^2)],\nonumber\\
\beta'' &=& -\frac{1}{r}\beta' - m^2\beta - 4mf\alpha\beta + 4f^2\beta(\alpha^2 + \beta^2) + g^{a\gamma} e B_{\textrm{\tiny B}} , \nonumber\\
A'' &=& -\frac{1}{r}A' - \frac{4g^{a\gamma}\alpha'(A' + A/r)}{1 - 4g^{a\gamma}\alpha},\nonumber\\
V''  &=& -\frac{1}{r}V' + 4g^{a\gamma}\frac{V'\alpha' - \beta'B_{\textrm{\tiny B}}}{1 - 4g^{a\gamma}\alpha}.\label{eqswesley}
\end{eqnarray}

\begin{figure}[H] 

    \begin{subfigure}{\linewidth} 
        \includegraphics[width=\linewidth]{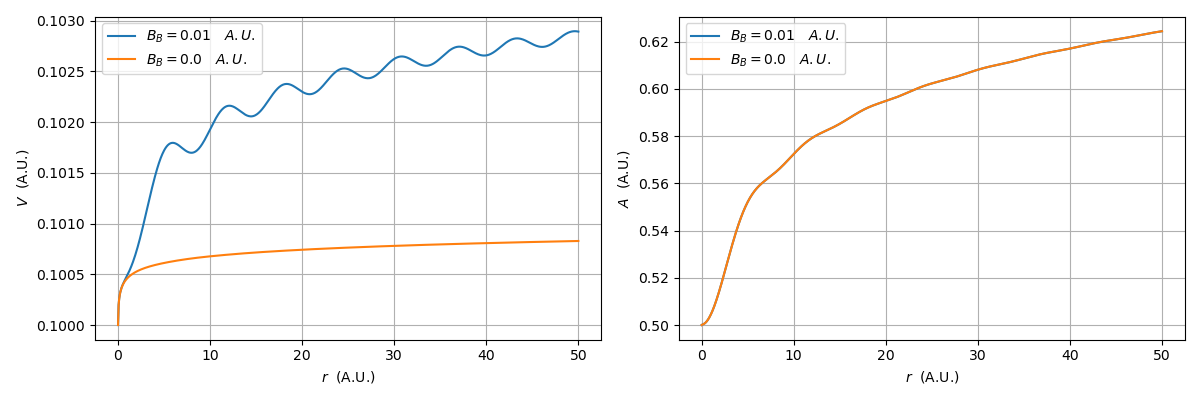}
        \caption{\justifying From the equation for the magnetic potential, $A$, in Eqs.~(\ref{eqswesley}), we observe that it does not depend on $B_{\textrm{\tiny B}}$. This explains the overlap of the blue and orange lines in the top-right graph.}
        \label{fig:1}
    \end{subfigure}

    \vspace{0.4cm}

    \begin{subfigure}{\linewidth} 
        \includegraphics[width=\linewidth]{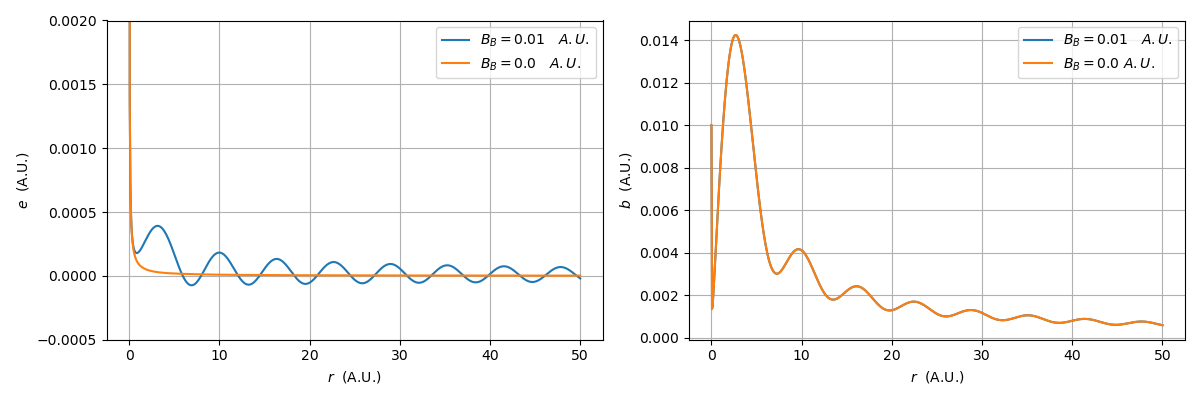}
        \caption{\justifying The magnetic field develops a configuration non-trivial at the origin, tending to zero away from the core. We suggest that the interaction between the axionic field, $\beta$, and the electromagnetic fields, $\boldsymbol{E}$ and $\boldsymbol{B}$, within the supersymmetric framework may be responsible for this behavior. A detailed study of these topological solutions will be presented in a forthcoming work.}
        \label{fig:2}
    \end{subfigure}

    \caption{\justifying Electric and magnetic potentials (a), and the corresponding electric and magnetic fields (b), as functions of $r$ in arbitrary units (A.U.).}
    \label{fig:combined}
\end{figure}
\begin{figure}[H] 

    \begin{subfigure}{\linewidth} 
        \includegraphics[width=\linewidth]{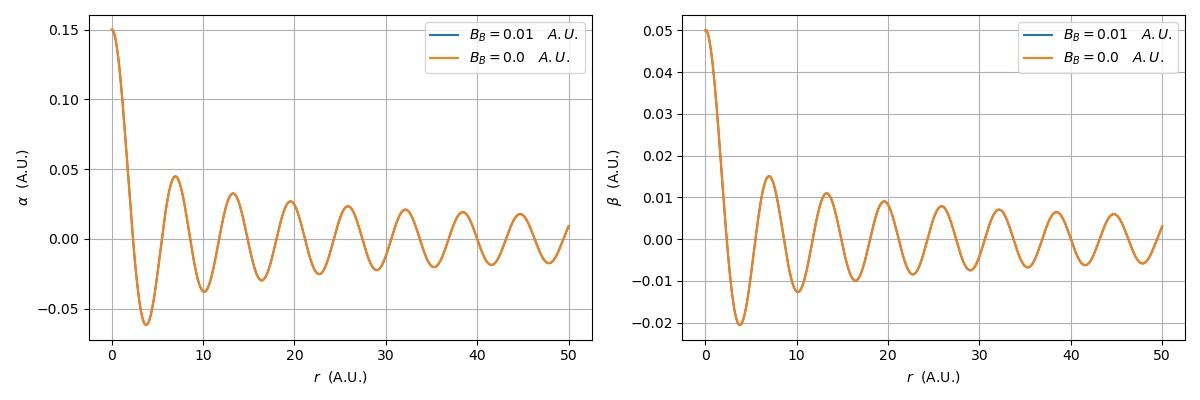}
        \caption{\justifying Low background magnetic field $B_{\textrm{\tiny B}}$. The blue and orange curves overlap due to the small  background magnetic field, revealing the fully coupled nonlinear dynamics of $\alpha$ and $\beta$, which oscillate in phase.
        }
        \label{fig:3}
    \end{subfigure}

    \vspace{0.4cm}

    \begin{subfigure}{\linewidth} 
        \includegraphics[width=\linewidth]{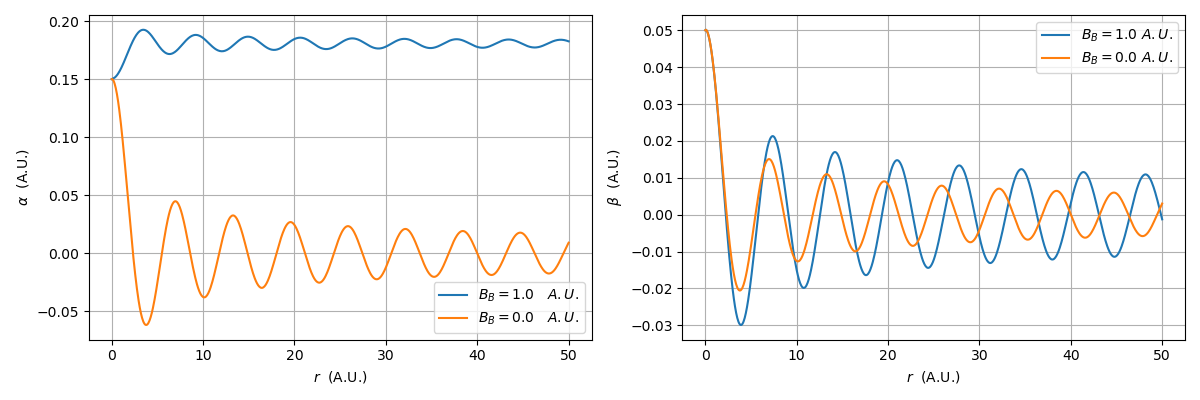}
        \caption{\justifying High background magnetic field $B_{\textrm{\tiny B}}=1$ (A.U.). In this case, the fields behave as if they were decoupled. This effect is enhanced for larger values of $g^{a\gamma}$ and $f$.
 As a result, in this case, measuring $\beta$ may be  easier.}
        \label{fig:4}
    \end{subfigure}
    \caption{\justifying The axion $\beta$ and the saxion $\alpha$  as a function of $r$ for different values of the background magnetic field.
$B_{\textrm{\tiny B}}$, in  arbitrary units (A.U.).   Note that the constant offset in $\alpha$ associated with its minimum
value has been subtracted in the plots. Working at the minimum, $\tilde{\alpha}=\alpha+m_a/f$, would correspond to a different topological configuration.  }
    \label{fig:axion}
\end{figure}

This constitutes a system of nonlinear differential equations. The corresponding numerical solutions are shown in Figs.~\ref{fig:1}--\ref{fig:4}. The different orders of magnitude of the terms in Eqs.~(\ref{eqswesley}) prevent a consistent graphical representation that simultaneously captures all contributions. Therefore, to qualitatively analyze the behavior of the interacting fields, we work in arbitrary units (A.U.). We fix the parameter values to $g^{a\gamma}=0.01\,$$\text{(A.U.)}$, $f=0.01$, $m_a=1.0\,$$\text{(A.U.)}$, with initial conditions $V_i =0.1$, $A_i=0.1$, $V'_i=0.01$, $A'_i=0.01$, $\alpha_i=0.15$, $\beta_i=0.05$, $\alpha'_i=0.01$, $\beta'_i=0.01$. Each graph shows two cases: the blue line corresponds to $B_{\textrm{\tiny B}}=0.01 \,\text{(A.U.)}$, while the orange line corresponds to a vanishing background field, $B_{\textrm{\tiny B}}=0$.





\section{Concluding remarks and future prospects}

The model presented and inspected throughout the previous Sections provides a supersymmetric realization of Axionic Electrodynamics. Accordingly, the resulting supersymmetric action describes the axion along with its supersymmetric partners, coupled to the electromagnetic sector. Quartic fermionic couplings and self-couplings, in addition to a non-polynomial interaction term, emerge as a consequence of supersymmetry. Such a landscape enriches Axionic Electrodynamics not only from a theoretical standpoint but also phenomenologically, as it provides additional interactions between possible dark matter candidates. 

We have computed the scalar potential and its minima, finding equal masses for the axion and its supersymmetric partners, leaving SUSY intact. However, when analyzing the DRs in the presence of an external magnetic field, we have found that the axion field acquires an effective mass, departing from the axion mass parameter  appearing in the original action. Furthermore, the supersymmetric structure of the model modifies Maxwell's equations by introducing fermionic contributions in the permittivity and permeability tensors of the constitutive relations. Some topological materials exhibit SUSY as an emergent phenomenon. The symmetry between fermion and boson fields can describe novel phase transitions and find characteristic critical exponents \cite{Chen_2013 ,Jian_2015,Sekine_2021}.  The effective structures generated by SUSY in our model may be viewed as a clue for emergent electromagnetic phenomena in systems that behave as Weyl materials. Therefore, similar effective electromagnetic responses with fermionic corrections to the constitutive equations deserve to be investigated in future work. We have also shown that SUSY modifications in Axionic Electrodynamics may lead to nontrivial configurations generated in the presence of background fields. We expect that such effects could be tested in a high-energy regime characterized by strong magnetic fields, for example,  high-energy astrophysical environments such as neutron stars. And it remains to be studied in depth the SUSY breaking triggered by the introduction of the magnetic background. 

Before concluding, we emphasize our interest in further  investigating  other implications of the action (\ref{accion4comp}). Several aspects of the model appear particularly promising and motivate further studies. For instance, the axion–axino–photino and quartic fermionic interaction terms in  (\ref{accion4comp}) suggest a Nambu–Jona-Lasinio-type interaction \cite{NAMBU1.122.345, nambu2.124.246},  which may point to a corresponding dynamical mass generation mechanism for the photino. This becomes more evident if we rewrite the term by applying the Fierz identity, taking into account that $\Psi$ and $\Lambda$ are Majorana fermions,
\begin{equation}
\frac{(g^{a\gamma})^2}{1 + 4g^{a\gamma} \alpha} (\bar{\Lambda} \Psi)^2 
\approx\frac{(g^{a\gamma})^2}{4}(1 - 4g^{a\gamma} \alpha) \Big[(\bar{\Lambda}\Lambda)(\bar{\Psi}\Psi) + (\bar{\Lambda}\gamma_5\Lambda)(\bar{\Psi}\gamma_5\Psi) + (\bar{\Lambda}\gamma^\mu\gamma_5\Lambda)(\bar{\Psi}\gamma_\mu\gamma_5\Psi)\Big] \,.
\end{equation}
The nonlinearity in $\alpha$ has been linearized by neglecting contributions beyond third order in $g^{a\gamma}$, as previously justified. By manipulating the quartic vertex, $(\bar{\Lambda} \Lambda)(\bar{\Psi} \Psi)$, a Nambu–Jona-Lasinio-type mechanism may occur, which could also have implications for the photino mass. Furthermore, the photino may also acquire a mass through another term in (\ref{accion4comp}), namely the quartic interaction $-\frac{1}{2}(g^{a\gamma})^2(\bar{\Lambda}\Lambda)^2$. We argue that a mechanism in which the four-fermion interaction leads to a possible non-vanishing vacuum expectation value of fermion condensates may result in a process of dynamical mass generation for the photino.

 Finally, let us focus our attention on the axion–photino–photon coupling term $g^{a\gamma}\sqrt{2}\,\bar{\Lambda} \Sigma^{\mu\nu} \Psi\, F_{\mu\nu}$, which also appears in  (\ref{accion4comp}). Based on its form, it could reasonably be interpreted as a  supersymmetric analog of the Primakoff effect. In the standard axion Primakoff effect \cite{sikivie}, an axion, $a$, converts into a photon $\gamma$ in the presence of an external electromagnetic field through the interaction
\begin{equation}
\mathcal{L}_{a\gamma} = -\frac{1}{4} g^{a\gamma} \beta\, F_{\mu\nu} \tilde{F}^{\mu\nu} = g^{a\gamma}\, \beta\, \boldsymbol{E} \cdot \boldsymbol{B}.
\end{equation}
In the supersymmetric model here described, a similar conversion may occur between the photino, $\Lambda$, and the axino, $\Psi$.  Analogously to what is done in the standard Primakoff effect, we need to expand the electromagnetic field into a background plus a propagating field,  \(F_{\mu\nu} = {F_{\textrm{\tiny B}}}_{\mu\nu}+ f_{\mu\nu}\), where \(
{F_{\textrm{\tiny B}}}_{0i} = {\boldsymbol E_{\textrm{\tiny B}}}_i, \; {F_{\textrm{\tiny B}}}_{ij} = -\varepsilon_{ijk} {\boldsymbol{B}_{\textrm{\tiny B}}}_k
\). The interaction term then becomes

\begin{align}
g^{a\gamma}\sqrt{2}\, (\bar{\Lambda} \Sigma^{\mu\nu} \Psi)\, F_{\mu\nu}
&= g^{a\gamma}\sqrt{2}\, (\bar{\Lambda} \Sigma^{\mu\nu} \Psi)\, f_{\mu\nu}
+ g^{a\gamma}\sqrt{2}\, (\bar\Lambda \Sigma^{0i}\Psi)\, {\boldsymbol E_{\textrm{\tiny B}}}_i
\nonumber \\
&\quad - 2 g^{a\gamma}\sqrt{2}\, \bar{\Lambda}
\left( \boldsymbol{\Sigma} \cdot \boldsymbol{B}_{\textrm{\tiny B}} \right) \Psi
\label{primakopf2}
\end{align}

Now, the production rate of photinos from axinos can be computed by considering a mixed propagator. To this end, the masses of the axino and the photino must first be determined using the methods discussed in previous Sections, allowing one to identify the heavier state and evaluate the decay rate into the lighter one.
Another interesting aspect appears when we examine the term $ g^{a\gamma}\sqrt{2}\,(\bar\Lambda \Sigma^{0i}\Psi)\,{\boldsymbol E_{\textrm{\tiny B}}}_i$ in (\ref{primakopf2}). This interaction could give rise to an Aharonov–Casher-type effect \cite{aharonov} if the axinos or photinos possess a nonzero magnetic moment. The Aharonov–Casher effect describes the interaction between a neutral particle with a magnetic moment and a background electric field. Similarly to the case of the anomalous magnetic moment of the electron or the magnetic moment of the neutrino \cite{giunti2015neutrinoelectromagneticinteractionswindow}, axinos and photinos,  although  electrically neutral, could acquire a magnetic moment through radiative corrections.  This illustrates how supersymmetric extensions may reproduce known phenomena while pointing to new particle sectors. We plan to focus on the possibilities raised here, some of which are currently under investigation in our collaboration.

\section*{Acknowledgments}

This work was financed in part by the Coordenação de Aperfeiçoamento de Pessoal de Nível Superior  (CAPES) and the Conselho Nacional de Desenvolvimento Científico e Tecnológico (CNPq). C.~Roldán-Domínguez also acknowledges the Centro Brasileiro de Pesquisas Físicas (CBPF) for the hospitality and for providing a stimulating research environment.

\section*{Data Availability Statement}

Data sharing not applicable to this article as no datasets were generated or analysed during the current study.



\end{document}